\documentclass[10pt,twoside,twocolumn,english,superscriptaddress,aps,preprint]{revtex4}
\usepackage[T1]{fontenc}
\usepackage[latin9]{inputenc}
\usepackage{color}
\usepackage{babel}
\usepackage{float}
\usepackage{textcomp}
\usepackage{amstext}
\usepackage{graphicx}
\usepackage[unicode=true,pdfusetitle,
 bookmarks=true,bookmarksnumbered=false,bookmarksopen=false,
 breaklinks=false,pdfborder={0 0 1},backref=false,colorlinks=false]
 {hyperref}

\makeatletter

\newcommand{\lyxmathsym}[1]{\ifmmode\begingroup\def\b@ld{bold}
  \text{\ifx\math@version\b@ld\bfseries\fi#1}\endgroup\else#1\fi}

\@ifundefined{textcolor}{}
{%
 \definecolor{BLACK}{gray}{0}
 \definecolor{WHITE}{gray}{1}
 \definecolor{RED}{rgb}{1,0,0}
 \definecolor{GREEN}{rgb}{0,1,0}
 \definecolor{BLUE}{rgb}{0,0,1}
 \definecolor{CYAN}{cmyk}{1,0,0,0}
 \definecolor{MAGENTA}{cmyk}{0,1,0,0}
 \definecolor{YELLOW}{cmyk}{0,0,1,0}
}

\usepackage{babel}
\usepackage{babel}

\makeatother

\begin{document}
\title{Unconventional magnetism in the 4\textit{d$^{4}$} based ($S=1$)
honeycomb system Ag$_{3}$LiRu$_{2}$O$_{6}$}
\author{R. Kumar}
\affiliation{Department of Physics, Indian Institute of Technology Bombay, Powai,
Mumbai 400076, India}
\author{Tusharkanti Dey}
\affiliation{Experimental Physics VI, Center for Electronic Correlations and Magnetism,
University of Augsburg, D-86159 Augsburg, Germany}
\author{P. M. Ette}
\affiliation{Central Electrochemical Research Institute-Madras Unit, CSIR-Madras
Complex, Taramani, Chennai 600113, India}
\author{K. Ramesha}
\affiliation{Central Electrochemical Research Institute-Madras Unit, CSIR-Madras
Complex, Taramani, Chennai 600113, India}
\author{Atasi Chakraborty}
\affiliation{Department of Solid State Physics, Indian Association for the Cultivation
of Science, Jadavpur, Kolkata 700032, India}
\author{I. Dasgupta}
\affiliation{Department of Solid State Physics, Indian Association for the Cultivation
of Science, Jadavpur, Kolkata 700032, India}
\author{J. C. Orain}
\affiliation{Laboratory for Muon Spin Spectroscopy, Paul Scherrer Insititut (PSI),
CH-5232 Villigen, Switzerland}
\author{C. Baines}
\affiliation{Laboratory for Muon Spin Spectroscopy, Paul Scherrer Insititut (PSI),
CH-5232 Villigen, Switzerland}
\author{Sándor Tóth}
\affiliation{Laboratory for Neutron Scattering and Imaging, Paul Scherrer Institute
(PSI), CH-5232 Villigen, Switzerland}
\author{A. Shahee}
\affiliation{Department of Physics, Indian Institute of Technology Bombay, Powai,
Mumbai 400076, India}
\author{S. Kundu}
\affiliation{Department of Physics, Indian Institute of Technology Bombay, Powai,
Mumbai 400076, India}
\author{M. Prinz-Zwick}
\affiliation{Experimental Physics V, Center for Electronic Correlations and Magnetism,
University of Augsburg, D-86159 Augsburg, Germany}
\author{A.A. Gippius}
\affiliation{Department of Physics, M.V. Lomonosov Moscow State University, 199991
Moscow, Russia}
\affiliation{P.N. Lebedev Physics Institute of Russian Academy of Science, 199991
Moscow, Russia}
\author{N. B{ü}ttgen }
\affiliation{Experimental Physics V, Center for Electronic Correlations and Magnetism,
University of Augsburg, D-86159 Augsburg, Germany}
\author{P. Gegenwart }
\affiliation{Experimental Physics VI, Center for Electronic Correlations and Magnetism,
University of Augsburg, D-86159 Augsburg, Germany}
\author{A.V. Mahajan}
\affiliation{Department of Physics, Indian Institute of Technology Bombay, Powai,
Mumbai 400076, India}
\date{\today}
\begin{abstract}
We have investigated the thermodynamic and local magnetic properties
of the Mott insulating system Ag$_{3}$LiRu$_{2}$O$_{6}$ containing
Ru$^{4+}$ (4$d$$^{4}$) for novel magnetism. The material crystallizes
in a monoclinic $C2/m$ structure with RuO$_{6}$ octahedra forming
an edge-shared two-dimensional honeycomb lattice with limited stacking
order along the $c$-direction. The large negative Curie-Weiss temperature
($\theta_{CW}=-57$~K) suggests antiferromagnetic interactions among
Ru$^{4+}$ ions though magnetic susceptibility and heat capacity show
no indication of magnetic long-range order down to 1.8~K and 0.4
K, respectively. \textcolor{black}{$^{7}$Li nuclear magnetic resonance
(NMR) shift follows the bulk susceptibility between 120-300~K and
levels off below 120~K. Together with a power-law behavior in the
temperature dependent spin-lattice relaxation rate between 0.2 and
2~K, it suggest dynamic spin correlations with gapless excitations.
Electronic structure calculations suggest an $S=1$ description of
the Ru-moments and the possible importance of further neighbour interactions
as also bi-quadratic and ring-exchange terms in determining the magnetic
properties. Analysis of our $\mu$SR data indicates spin freezing
below 5 K but the spins remain on the borderline between static and
dynamic magnetism even at 20 mK.}
\end{abstract}
\maketitle

\section{introduction}

Over the past few years, there has been a shift in focus from 3\textit{d}-based
systems to the exploration of 4\textit{d} and 5\textit{d}-based ones
due to the possibility of strong spin-orbit coupling (SOC) driving
exotic magnetism \cite{Okamoto2007,Kim2009,Singh2012,Choi2012,Takayama2015,Dey.2012,Dey.2013,Dey2013b,Borzi2007,Wu2011,Devries2010,Kumar2016,Krempa2014,Thompson2016,Taylor2016}.
The SOC is found to be very strong for 5\textit{d-}based systems and
could stabilize a Mott insulating state as also other novel phases
\cite{Okamoto2007,Kim2009,Dey.2012,Kumar2016,Krempa2014,Thompson2016,Taylor2016}.
Issues such as the realization of the $J_{eff}$ = 1/2 state for the
$d^{5}$ configuration (half-filled) \cite{Kim2009}, possible realization
of the Kitaev model \cite{Kitaev2006} in $d^{5}$ Mott insulators
\cite{Takayama2015,Jackeli2009}, and the emergence of spin-liquid
states in triangular lattice materials have been widely explored for
Ir-based materials \cite{Dey.2012,Dey.2013,Kumar2016,Catuneanu2015,Becker2015}.

\begin{figure}[t]
\begin{centering}
\includegraphics[clip,scale=0.35]{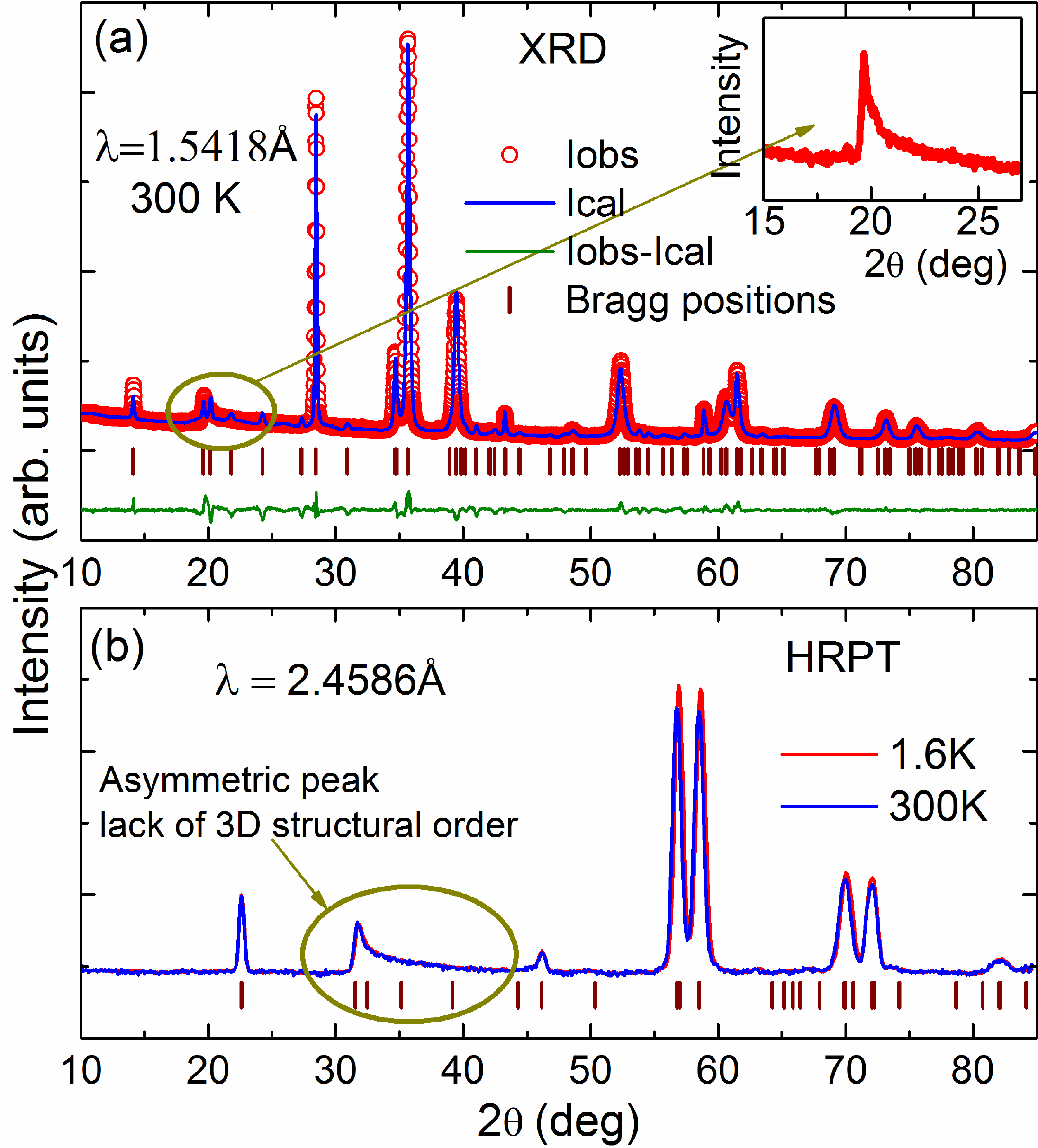} 
\par\end{centering}
\caption{\label{XRD} (a) X-ray diffraction data collected for a powder sample
of Ag$_{3}$LiRu$_{2}$O$_{6}$ at 300~K. Inset depicts the asymmetric
peak (a characteristic of 2D structural order) around $20\lyxmathsym{\protect\textdegree}$.
(b) Neutron diffraction data at 300~K (violet solid line) and 1.6~K
(cyan solid line) with $\lambda=2.4586$ Å. Encircled peak (dark yellow)
at Bragg angle $31.5\lyxmathsym{\protect\textdegree}$ is the asymmetric
peak.}
\end{figure}
However, a very interesting scenario could arise for materials away
from half-filling; such as four electrons in the $t_{2g}$ manifold.
It has been proposed by Khaliullin \cite{Khaliullin2013} that for
oxide systems containing Ru$^{4+}$, Re$^{3+}$, Os$^{4+}$ or Ir$^{5+}$,
there can often be comparable values of SOC ($\lambda\sim50-200$~meV)
and superexchange energy scales ($\frac{4t^{2}}{U}\sim50-100$~meV),
which could give rise to excitonic magnetism and resultant novel phases.
Recently, Meetei \textit{et al}. \cite{Meetei2015} and Svoboda \textit{et
al}. \cite{Svoboda2017} have worked further on this and suggested
the possible formation of a spin-orbital liquid even in the absence
of geometric frustration. Further, theoretical/experimental attempts
have been made to realize this novel magnetism in Ir$^{5+}$ based
perovskite NaIrO$_{3}$ \cite{Bremholm2011}, double perovskites Ba$_{2}$YIrO$_{6}$
and Sr$_{2}$YIrO$_{6}$ \cite{Dey2016,Hammerath2017,Chen2017,Cao2014,Corredor2017,Bhowal2015,Pajskr2016}
and triple perovskite Ba$_{3}$ZnIr$_{2}$O$_{9}$ \cite{Nag2016}.
However, conclusive evidence of this novel magnetism is still elusive.
This implies that one should explore materials with a lower SOC and
Ru$^{4+}$ materials might be a good starting point. Recent theoretical
studies also proposed Ru-based materials as good candidates to search
for excitonic magnetism \cite{Meetei2015,Svoboda2017}.

In this report, we detail the structural, bulk, and local magnetic
properties of a \textit{t$^{4}{}_{2g}$} based honeycomb system Ag$_{3}$LiRu$_{2}$O$_{6}$
\cite{Kimber2010} using x-ray diffraction, neutron diffraction, heat
capacity, muon spin rotation ($\mu$SR) and nuclear magnetic resonance
(NMR) techniques. The honeycomb structure decorated with any of the
$d^{4}$ ions (Ru$^{4+}$, Re$^{3+}$, Os$^{4+}$, and Ir$^{5+}$)
has been proposed to manifest novel physical properties \cite{Khaliullin2013}.
In Ag$_{3}$LiRu$_{2}$O$_{6}$, structurally, Ru atoms give rise
to a honeycomb geometry with the Li atom sitting at the center of
the honeycomb. Our bulk data do not show any magnetic ordering down
to 1.6\,K inspite of strong antiferromagnetic interactions. A magnetic
contribution to the specific heat is present compared to the non-magnetic
analog Ag$_{3}$LiTi$_{2}$O$_{6}$, however, without any sharp anomaly.
Static susceptibility deduced from the $^{7}$Li-NMR line shift shows
a plateau below 120\,K and down to 150\,mK. These signatures suggest
that the static spin correlations are absent/frozen out in the material.
The $^{7}$Li nuclear spin-lattice relaxation rate 1/$T$$_{1}$ decreases
with decreasing temperature $T$ without any anomaly and displays
a power law ($T^{4}$) behaviour below 2\,K. This is suggestive of
magnetic moments remaining dynamic and the excitations being gapless.
From our $\mu$SR measurements, Ag$_{3}$LiRu$_{2}$O$_{9}$ presents
a spin-glass-like ground state with a transition temperature $T_{g}$=5.5(5)~K
though the spins display behaviour which is at the borderline between
static and dynamic even at 20 mK. Our \textit{ab-initio} electronic
structure calculations infer negligible SOC and point towards a ferromagnetic
coupling between the three nearest neighbours of each Ru and antiferromagnetic
further neighbor couplings. This, coupled with deviations from the
Heisenberg model, possibly results in frustration which might drive
the observed behavior.

\section{experimental details}

The polycrystalline samples of Ag$_{3}$LiRu$_{2}$O$_{6}$ and Ag$_{3}$LiTi$_{2}$O$_{6}$
(nonmagnetic analog used for heat capacity analysis) were prepared
in two steps. The precursor Li$_{2}$RuO$_{3}$ was synthesized by
the solid state reaction route by firing stoichiometric amounts of
Li$_{2}$CO$_{3}$ and Ru at $1000$$^{\circ}$C for $12$ hours in
an alumina crucible and followed by another heating cycle at 950$^{\circ}$C
for 24 hours after grinding the sample and mixing $10$\% excess Li$_{2}$CO$_{3}$.
The nonmagnetic Li$_{2}$TiO$_{3}$ was prepared by firing a stoichiometric
mixture of Li$_{2}$CO$_{3}$ and TiO$_{2}$ at $1000$$^{\circ}$C.
Having obtained single phase samples of the precursors Li$_{2}$RuO$_{3}$
and Li$_{2}$TiO$_{3}$, high purity AgNO$_{3}$ was mixed with each
of the starting materials in the ratio $1:10$ to prepare the final
compositions of Ag$_{3}$LiRu$_{2}$O$_{6}$, and Ag$_{3}$LiTi$_{2}$O$_{6}$.
The crucibles containing the mixture of materials in a $1:10$ ratio
were slowly heated to $300$$^{\circ}$C in air and held at this temperature
for $6$ hours followed by cooling to room temperature. The residual
AgNO$_{3}$ and the reaction byproduct LiNO$_{3}$ were removed by
washing the materials with water. X-ray diffraction measurements on
the powder samples at room temperature were performed with a Panalytical
Xpert Pro diffractometer using Cu-$\mathrm{K_{\alpha}}$ radiation.
Neutron diffraction data were taken on the HRPT beamline at the Paul
Scherrer Institute PSI at 300~K and 1.6~K using a wavelength $\lambda=2.4586$
Å.

Magnetisation $M$ measurements as a function of applied field $H$
(0 to 90 kOe) and temperature $T$ (in the range 1.8 K to 400 K))
were performed using a Quantum Design SQUID VSM. Zero-field cooled
(ZFC) and field cooled (FC) magnetisation measurements in a low field
of 25 Oe were performed down to 1.8 K. The heat capacity $C_{\mathrm{p}}(T)$
was measured with a Quantum Design PPMS in various applied fields
down to about 0.4\,K. The $\mu$SR experiments were performed on
a powder sample at the PSI. In the high temperature range (1.5~K
< $T$ < 200~K) we used the General Purpose Surface-muons instrument
(GPS). We mounted about 1~g of sample in a 15~mm$\times$15~mm
Al envelope on a Cu fork. Therefore the sample stops the whole muon
beam and we can neglect the experimental background. For the low temperature
regime (20~mK < $T$ < 17.5~K) we used the Low Temperature Facility
Instrument (LTF). We glued about 1~g of powder with GE-varnish on
a silver plate to ensure thermal conductivity. Additionally, local
probe nuclear magnetic resonance NMR measurements were performed on
the $^{7}$Li nucleus in a fixed field of 93.95 kOe as also at a fixed
frequency of $95$\,MHz. The variation of lineshape with $T$ was
measured as also that of the spin-lattice relaxation rate 1/$T_{1}$
down to 150 mK. 

\section{results}

\subsection{Xrd and structural details}

\begin{figure}[t]
\begin{centering}
\includegraphics[bb=5bp 270bp 610bp 550bp,clip,scale=0.39]{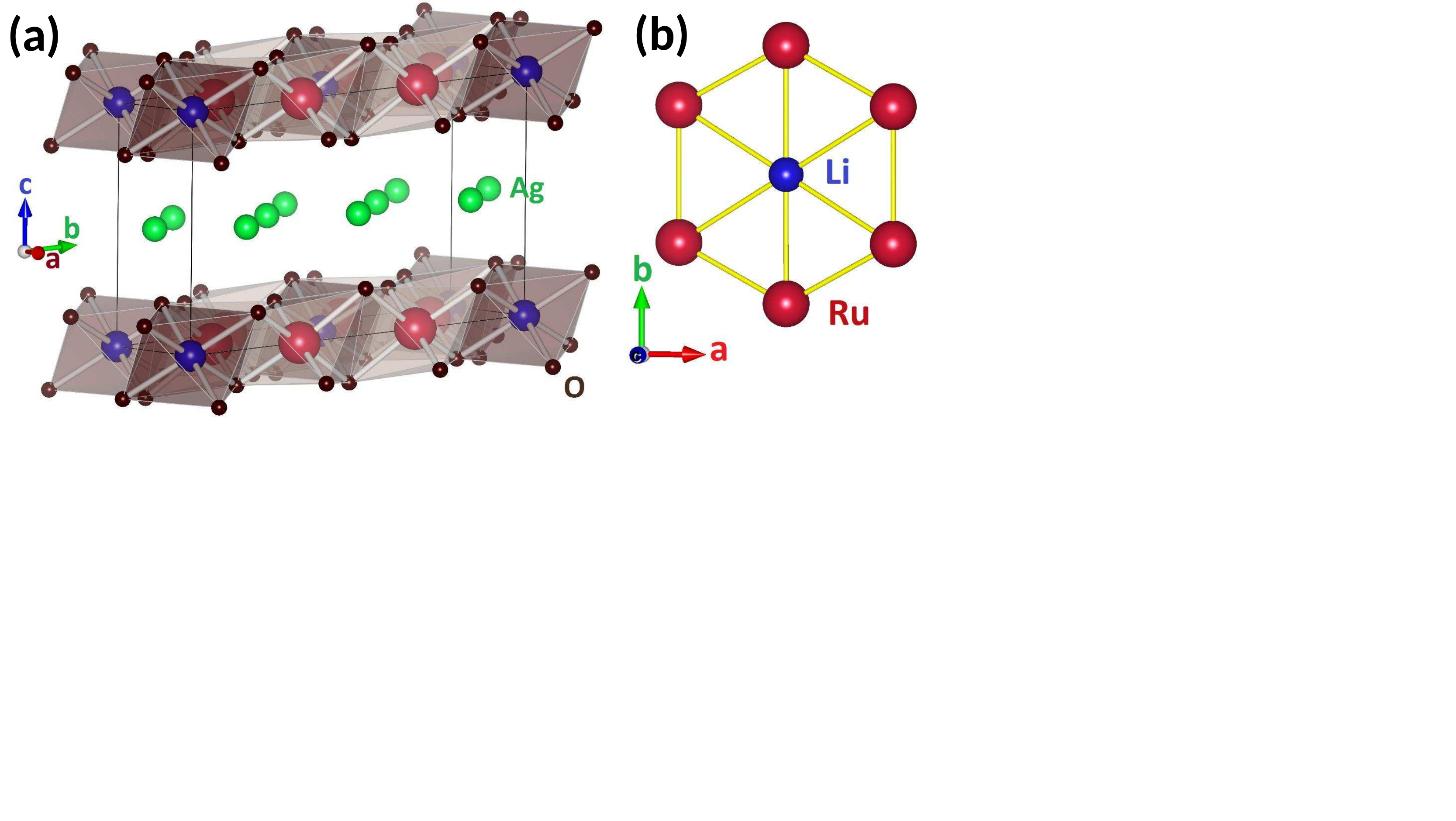} 
\par\end{centering}
\caption{\label{Structure} (a) A unit cell of Ag$_{3}$LiRu$_{2}$O$_{6}$
with RuO$_{6}$ (Ru shown as pink balls) and LiO$_{6}$ (Li shown
as blue balls) octahedra in the crystallographic $a-b$ plane. (b)
A depiction of 2D edge sharing honeycomb lattice formed by Ru atoms
in the $a-b$ plane with Li atom sitting at the center of the honeycomb. }
\end{figure}

The diffraction patterns in Fig. \ref{XRD} show a sawtooth shaped
peak at low angles (see inset of Fig. \ref{XRD}(a) for x-ray data
and Fig. \ref{XRD}(b) for neutron diffraction data) which is commonly
known as the Warren peak and is characteristic of 2D structural order
with stacking faults in the $c$-direction \cite{Warren1941}. Note
that stacking faults are not uncommon in such systems, for instance,
in Na$_{2}$IrO$_{3}$ (Ref. \cite{Choi2012}), in Li$_{2}$RhO$_{3}$
(Ref. \cite{Khuntia2017}) as also in $\alpha$-RuCl$_{3}$ (Ref.
\cite{Majumder2015}). This is not likely to affect the two-dimensional
magnetic properties. The neutron diffraction data do not evidence
the appearance of additional Bragg peaks down to 1.6 K. Further, absence
of magnetic long-range order in Ag$_{3}$LiRu$_{2}$O$_{6}$ is not
due to stacking faults as we have, in fact, observed LRO in the structurally
analogous Ag$_{3}$LiMn$_{2}$O$_{6}$ (Ref. \cite{Kumar-un}). The
x-ray diffraction pattern of Ag$_{3}$LiRu$_{2}$O$_{6}$ could be
successfully indexed with the monoclinic structure under space group:
$C\,2/m$ (Space Group no. 12), $Z=2$, and the Rietveld refinement
of the x-ray diffraction data with the FULLPROF suite \cite{Carvajal1993}
yields the profile parameters $R_{wp}=4.66$$\%$, $R_{exp}=2.72$$\%$,
$R_{p}=3.46$$\%$ and $\chi^{2}=2.93$. The obtained lattice parameters
$a=5.2248(9)$ $\textrm{Å}$, $b=9.0459(15)$ $\textrm{Å}$, $c=6.5101(12)$
$\textrm{Å}$ and $\beta=74.480(12)\lyxmathsym{\textdegree}$ are
in excellent agreement with the previously reported results \cite{Kimber2010}.\textbf{
}In Ag$_{3}$LiRu$_{2}$O$_{6}$, Ru/Li ions coordinate with eight
surrounding oxygen atoms and make RuO$_{6}$/LiO$_{6}$ octahedra
(see Fig. \ref{Structure} a). The RuO$_{6}$ octahedra connect in
an edge-sharing fashion and give rise to a honeycomb network and the
Ru ions are best viewed as forming a two-dimensional (2D) honeycomb
lattice in the $a-b$ plane (see Fig. \ref{Structure} b). Note that
although there is a unique Ru site, there are inequivalent O sites.
As a result, there are two types of Ru-O-Ru bonds between a Ru and
its three nearest neighbour Ru. As seen later, this results in two
different couplings between a Ru and its nearest neighbours. The incorporation
of Ag atoms into the primary material Li$_{2}$RuO$_{3}$ actually
works as an intercalation between the Ru layers and essentially makes
Ag$_{3}$LiRu$_{2}$O$_{6}$ a 2D system.

\subsection{Magnetisation}

\begin{figure}[h]
\centering{}\includegraphics[clip,scale=0.35]{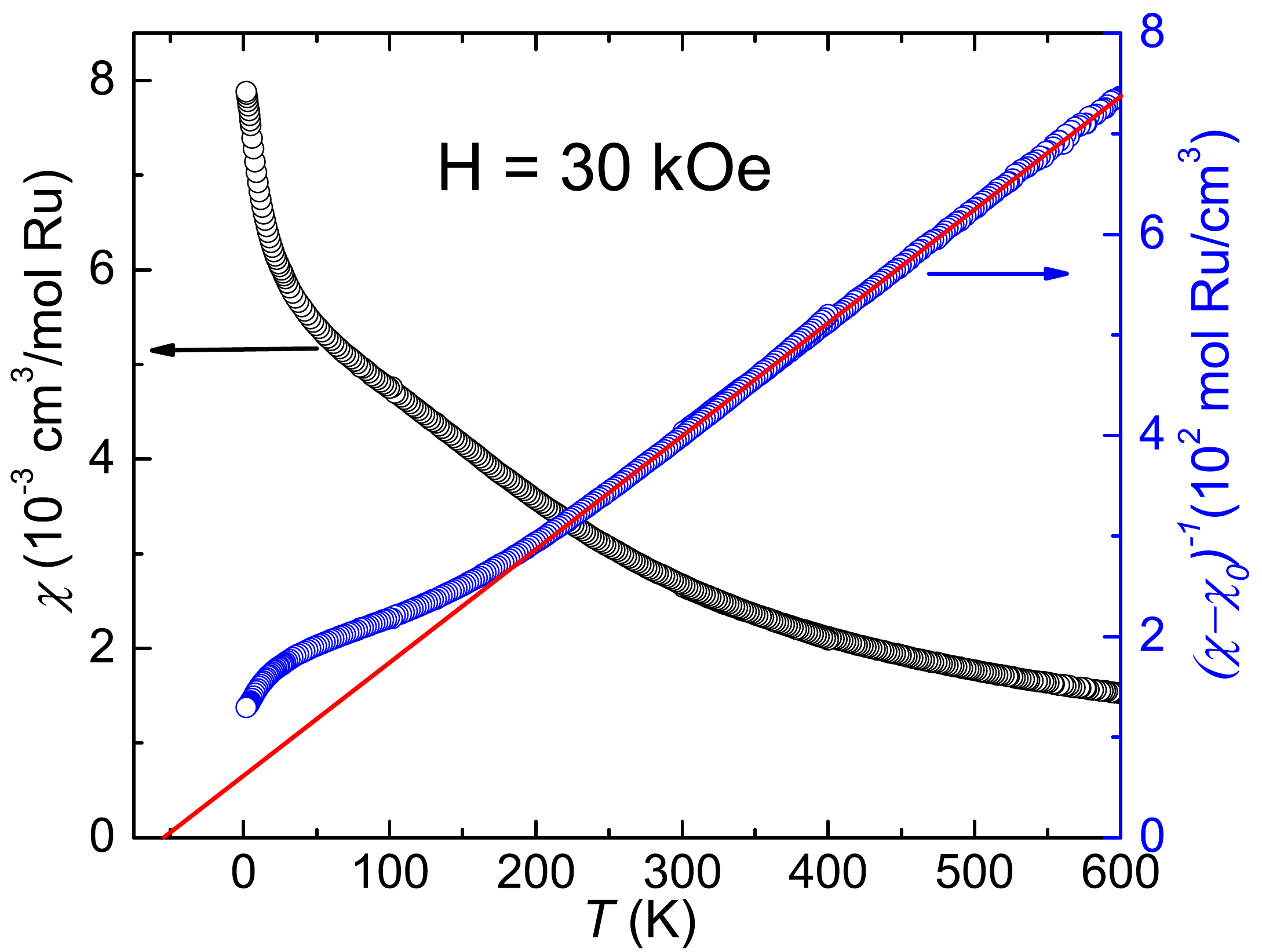}
\caption{\label{susceptibility} (a) Variation of susceptibility (left \textit{y}-axis)
and inverse susceptibility (right \textit{y}-axis) for Ag$_{3}$LiRu$_{2}$O$_{6}$
with $T$ in the range 2-600~K. The dotted line through the inverse
susceptibility data intercepts the temperature axis aound $-57$\,K.}
\end{figure}

\begin{figure}
\includegraphics[scale=0.3]{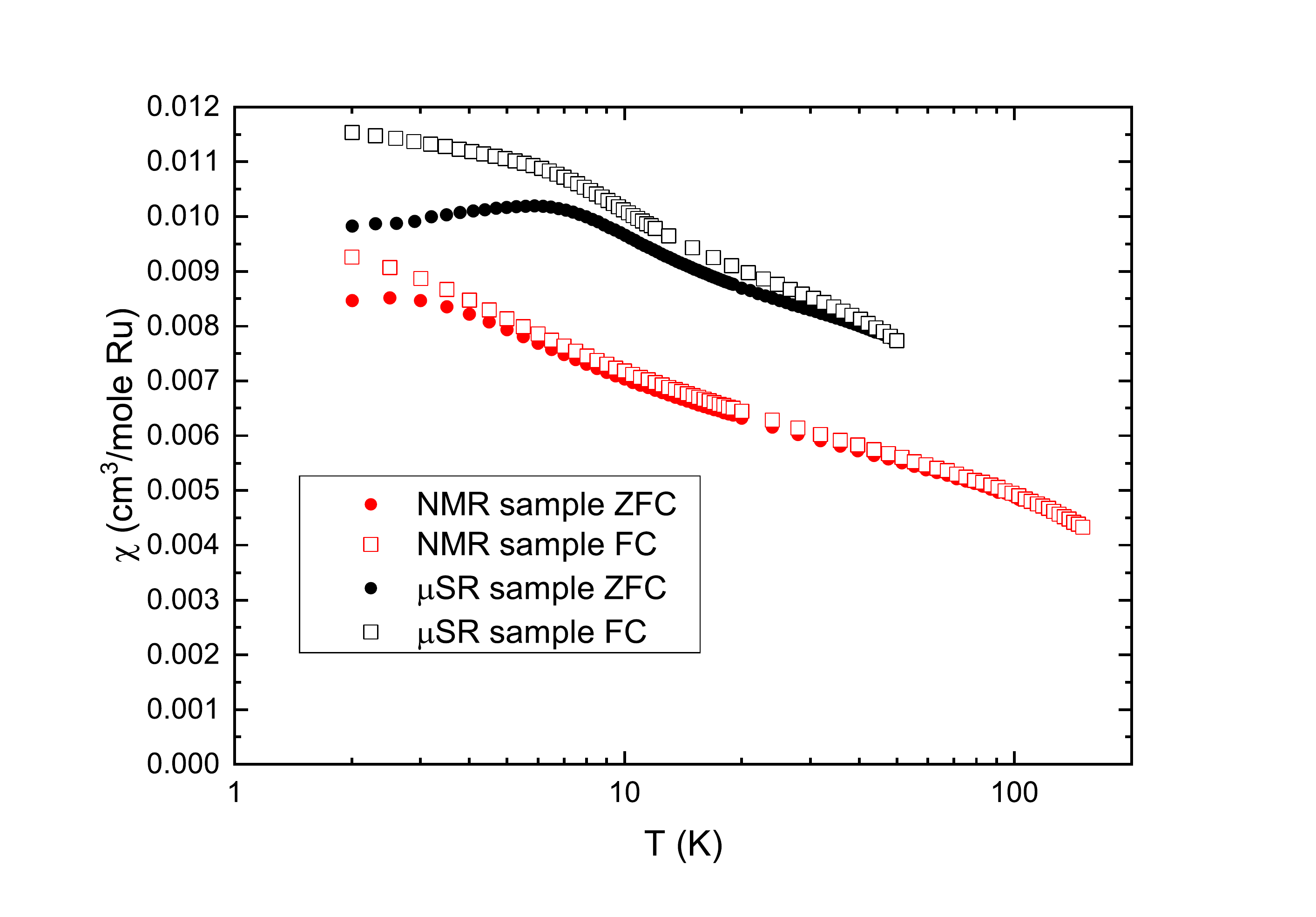}\caption{\label{fig:-zfcfc}$\chi=M/H$ measured in an applied field of 25
Oe is shown as a function of $T$. ``NMR sample'' refers to the
one on which all the measurements were done. Neutron diffraction and
$\mu$SR measurements were done on the ``$\mu$SR sample''.}
\end{figure}

Figure \ref{susceptibility} shows the DC susceptibility $\chi(T)$
of Ag$_{3}$LiRu$_{2}$O$_{6}$ measured in the $T$-range $2-600$~K
on a Quantum Design MPMS with the oven option. The $\chi(T)$ data
do not exhibit any anomaly in the $T$-range $2-600$~K though there
are hints of a plateau around 100 K. Our neutron diffraction data,
see Fig. \ref{XRD}(b), collected down to $1.6$~K with wavelength
$\lambda=2.4586$ Å do not show any evidence of a phase transition
either. A fit of the data to the Curie-Weiss law ($\chi=\chi_{0}+\frac{C}{T-\theta_{CW}}$)
in the $T$-range $300-600$~K gives $\chi_{0}=1.7\times10^{-4}$cm$^{3}$/mol
Ru and the asymptotic Curie-Weiss temperature $\theta_{CW}=-57$~K.
The negative $\theta_{CW}$ infers the presence of antiferromagnetic
coupling between Ru moments. Note that if $\chi_{0}$ is not left
as a free parameter but fixed to a larger value, it yields a smaller
$\theta_{CW}$ and a smaller Curie constant, though with a poorer
fit. Measurements to even higher temperatures would have helped obtain
$\chi_{0}$ with better accuracy but the sample degrades at higher
temperatures. The value of the Curie constant $C$ is about 0.88 cm$^{3}$
K/mol Ru. This leads to an effective moment of 2.65 $\mu_{B}$ which
is slightly smaller than the expected spin-only value (for $S$ =
1) of $2.83$ $\mu_{B}$. Magnetisation under zero field cooled (ZFC)
and field cooled (FC) conditions was measured in a low field of 25
Oe. This is shown in Fig. \ref{fig:-zfcfc}. We find that there is
a weak ZFC-FC bifurcation below about 3K. In another sample of Ag$_{3}$LiRu$_{2}$O$_{6}$
(from a different batch) on which detailed $\mu$SR measurements (as
also neutron diffraction) were performed, the bifurcation is greater
as also at a higher temperature of about 6 K. This could arise from
a fraction of moments in the sample (extrinsic or intrinsic) which
freeze. 


\subsection{Heat capacity}

\begin{figure}[h]
\centering{}\includegraphics[clip,scale=0.32]{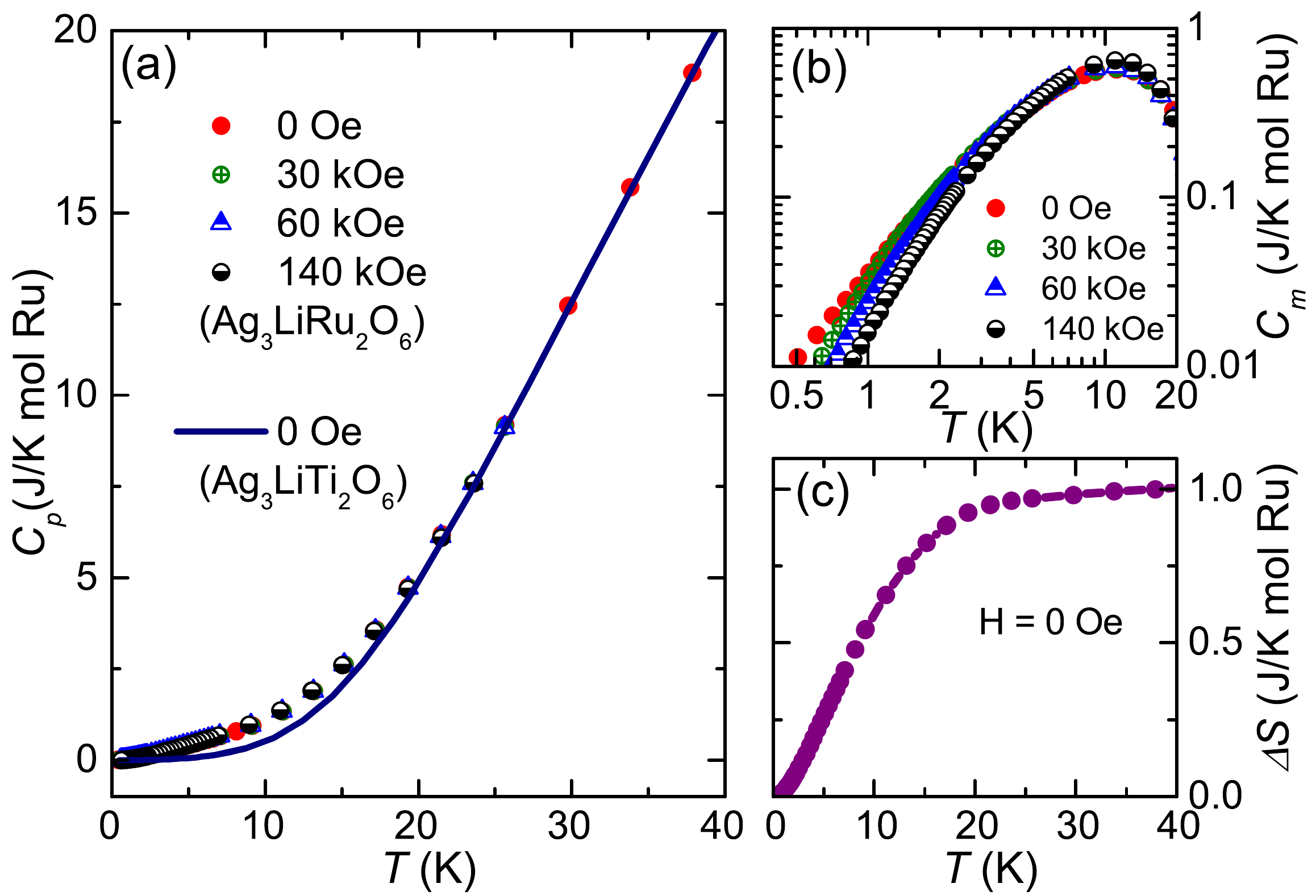}
\caption{\label{HeatCapacity} (a) Heat capacity for Ag$_{3}$LiRu$_{2}$O$_{6}$
at different fields (various symbols) and for non-magnetic analog
Ag$_{3}$LiTi$_{2}$O$_{6}$ at zero field (solid line) are shown
as function of temperature. (b) Magnetic specific heat $C$$_{m}$
as a function of $T$ for Ag$_{3}$LiRu$_{2}$O$_{6}$ with $H$ =
0, 30, 60, and 140 kOe. (d) Entropy change at $H$ = 0 Oe for Ag$_{3}$LiRu$_{2}$O$_{6}$.}
\end{figure}

Heat capacity measurements were made to probe low-energy excitations
associated with possible magnetism in the sample. Whereas data were
taken in a larger temperature range, Fig. \ref{HeatCapacity}(a) depicts
the heat capacity data ($C$$_{p}$) of Ag$_{3}$LiRu$_{2}$O$_{6}$
and structurally identical (nonmagnetic) Ag$_{3}$LiTi$_{2}$O$_{6}$
in the temperature range $0.4-40$~K. The measured heat capacity
for Ag$_{3}$LiRu$_{2}$O$_{6}$ does not show any significant dependence
on the magnetic field and remains featureless in the measured $T$-range.
To extract the magnetic specific heat $C_{m}$ of Ag$_{3}$LiRu$_{2}$O$_{6}$
a procedure as in Ref. \cite{Bouvier1991} was employed. The heat
capacity of nonmagnetic Ag$_{3}$LiTi$_{2}$O$_{6}$ was measured.
The ratio of the Debye temperatures of the Ru-compound and the Ti-compound
$\frac{\theta_{D}(Ru)}{\theta_{D}(Ti)}$ was determined using the
procedure of Ref. \cite{Bouvier1991}. The temperature axis of the
Ag$_{3}$LiTi$_{2}$O$_{6}$ was multiplied by the ratio of the Debye
temperatures ($\frac{\theta_{D}(Ru)}{\theta_{D}(Ti)}$ = 0.95) before
the specific heat of Ag$_{3}$LiTi$_{2}$O$_{6}$ was subtracted from
the total specific heat of Ag$_{3}$LiRu$_{2}$O$_{6}$. The magnetic
specific heat $C_{m}$ thus obtained is shown in Fig. \ref{HeatCapacity}(b).
At low-$T$, a power law behaviour is seen with an exponent of about
$1.65$. The calculated entropy change ($\triangle S$) for Ag$_{3}$LiRu$_{2}$O$_{6}$,
shown in Fig. \ref{HeatCapacity}(c), is estimated to be about $11\%$
of $9.12$\,J/K mol Ru expected for $S=1$. The power law $T$-dependence
of $C_{m}$ and the large quenching of $\triangle S$ supports the
realization of a highly degenerate ground state which is presumably
gapless.

\subsection{$^{7}$Li NMR}

\begin{figure}[h]
\begin{centering}
\includegraphics[bb=0cm 0cm 25cm 22cm,clip,scale=0.28]{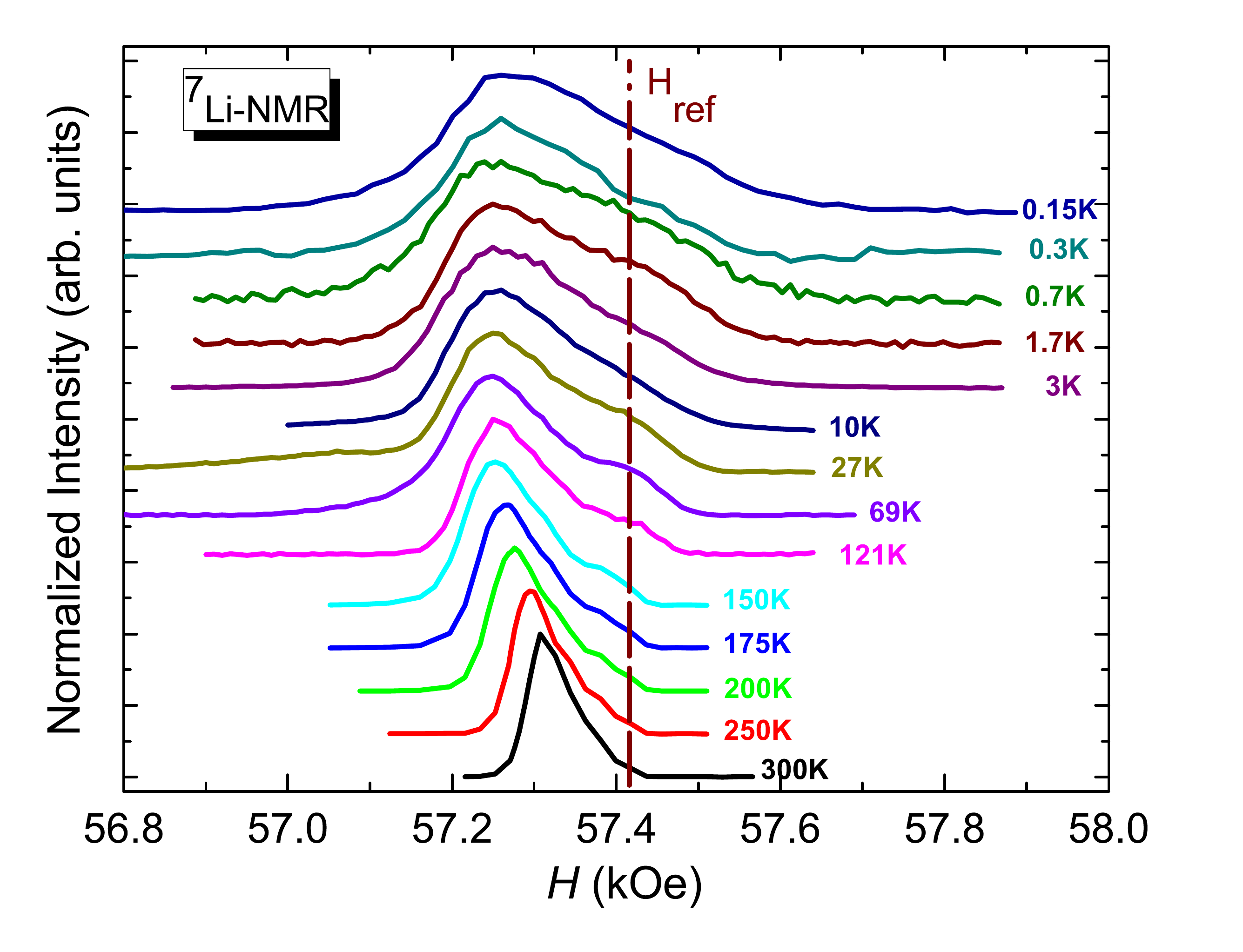} 
\par\end{centering}
\caption{\label{spectra} $T$-dependence of $^{7}$Li-NMR (normalized) spectra
for Ag$_{3}$LiRu$_{2}$O$_{6}$ is shown. The shift is seen to be
temperature independent from $120$\,K down to about $150$\,mK.
Note that the spectra above $121$\,K were actually obtained from
the echo integral at various frequencies in a fixed field of $93.94$\,kOe.
For these spectra, the $x$-axis was then converted to field units
to compare with the lower temperature data which were obtained by
sweeping the field at a fixed frequency of $95$ MHz.}
\end{figure}
\begin{figure}
\begin{centering}
\includegraphics[scale=0.37]{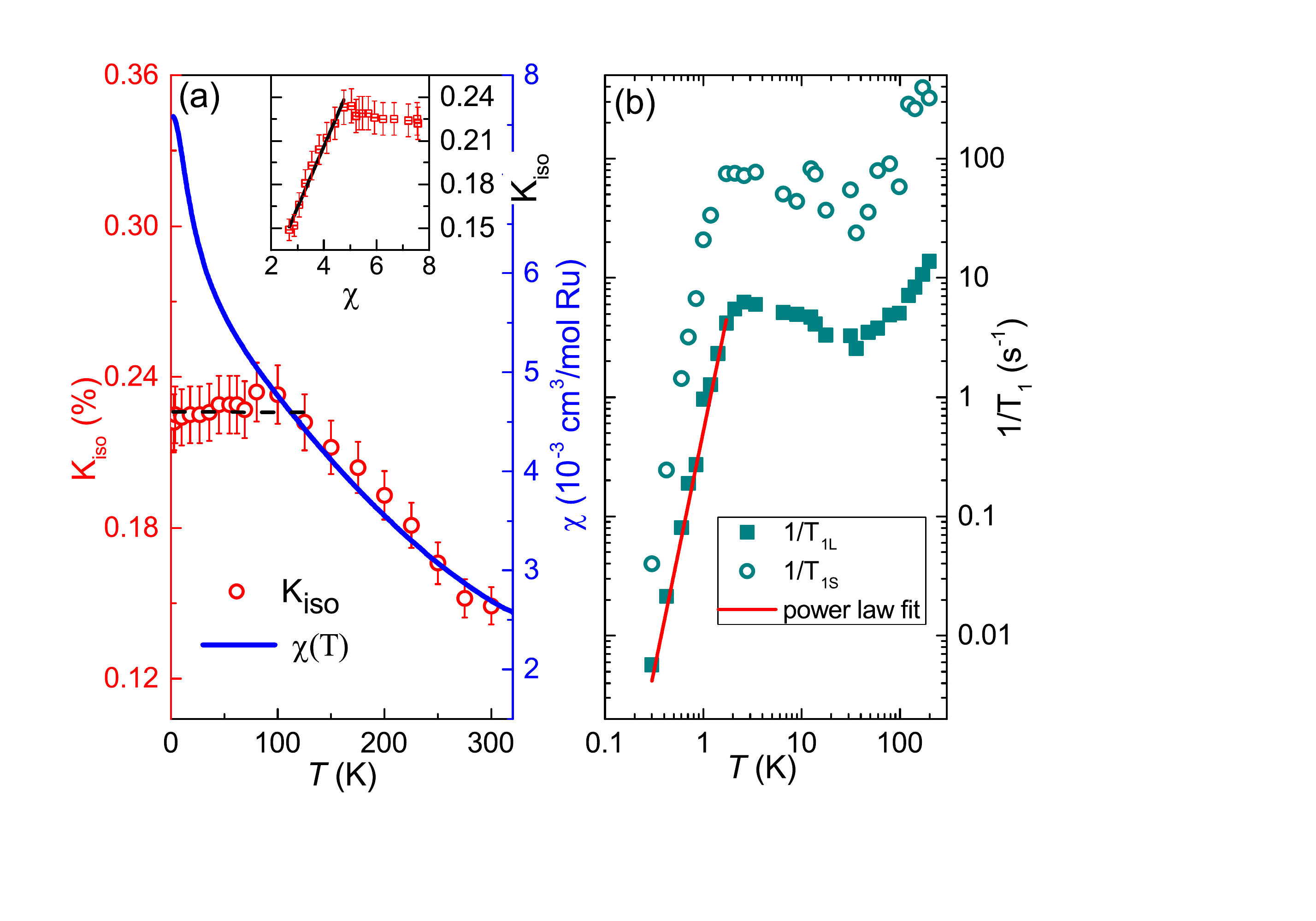} 
\par\end{centering}
\caption{\label{Kiso and T1} (a) $^{7}$Li-NMR line shift ($K_{iso}$) and
the bulk susceptibility ($\chi$) are plotted as a function of temperature,
on the left and right axes, respectively. The inset illustrates the
$^{7}$Li-NMR line shift plotted against the bulk magnetic susceptibility
($\chi$) with temperature as an implicit parameter. (b) $^{7}$Li
nuclear spin-lattice relaxation rate plotted against temperature.
The solid line is a power-law fit of the $1/T_{1L}$ data yielding
a $T^{4}$ variation. }
\end{figure}

NMR, being a local probe, is instrumental\textbf{ }in identifying
the change in magnetization at a local level. We performed $^{7}$Li
(nuclear spin: $I=\frac{3}{2}$, gyromagnetic ratio: $\frac{\gamma}{2\pi}=16.546\,$MHz/T)
NMR spectra measurements (echo integral at variable frequency) at
various temperatures from $300$\,K down to $150$~mK. Data were
obtained in two ways: (i) in a fixed field of $H=93.954$~kOe, the
echo integral was obtained as a function of frequency in the temperature
range $300$\,K to $80$\,K and (ii) at a fixed frequency of $95$\,MHz,
the echo integral was obtained as a function of the field in the $T$-range
$120$\,K to $150$~mK. The spectra are displayed together in Fig.
\ref{spectra} after scaling the $x$-axis of the frequency sweep
data with the gyromagnetic ratio of $^{7}$Li to obtain it in field
units corresponding to the frequency of the field sweep measurements.
The $^{7}$Li-NMR spectra were found to be asymmetric throughout the
measured temperature range: (\textit{i}) a main peak, which qualitatively
displays a variation with temperature and (\textit{ii}) a shoulder
on the higher field side which is centered around the zero-shift position,
and which remains almost unaffected in the entire temperature range
(see Fig. \ref{spectra}). The appearance of this shoulder in $^{7}$Li-NMR
powder spectra is most likely due to the anisotropy of the hyperfine
field. Any significant Li/Ru antisite structural disorder is ruled
out by our x-ray analysis. In the light of this anisotropy, the static
susceptibility was estimated by extracting the $K_{iso}$ (powder
averaged line shift) as a function of $T$ by matching the experimental
$^{7}$Li-NMR spectra with the simulated one. The $K_{iso}$ follows
the bulk susceptibility data, suggestive of a significant hyperfine
coupling between Ru and Li atoms. The $K_{iso}$ becomes nearly $T$-independent
(or perhaps weakly decreases) below about $120$\,K as seen in Fig.
\ref{Kiso and T1}(a). So the low-temperature rise in the bulk susceptibility
appears to be driven by some extrinsic Curie contributions. A finite
shift at the lowest temperature suggests gapless spin excitations.
Note however that NMR measurements are made in a magnetic field and
a finite shift could result from the closing of the gap due to the
field but the zero-field heat capacity data exclude the possibility
of a gap. The width of the spectrum remains unchanged below $T\approx2$
K (see Fig. \ref{spectra}). The hyperfine coupling constant $A_{hf}$
follows the relation: $K_{iso}=K_{chem}+\frac{A_{hf}}{N_{A}\mu_{B}}\chi_{spin}$
($K_{chem}$ is the chemical shift, $N_{A}$ is the Avogadro number,
$\mu_{B}$ is the Bohr magneton and $\chi_{spin}$ is the bulk susceptibility).
From the slope of $K_{iso}$ vs $\chi_{spin}$ plot $A_{hf}$ was
found to be 2.34(13) kOe/$\mu_{B}$ with $K_{chem}=0.03(1)$\%, as
shown in the inset of Fig. \ref{Kiso and T1}(a). The value of the
hyperfine coupling will turn out somewhat larger in case the fitting
range is limited to higher temperatures. 

\begin{figure}
\begin{centering}
\includegraphics[scale=0.3]{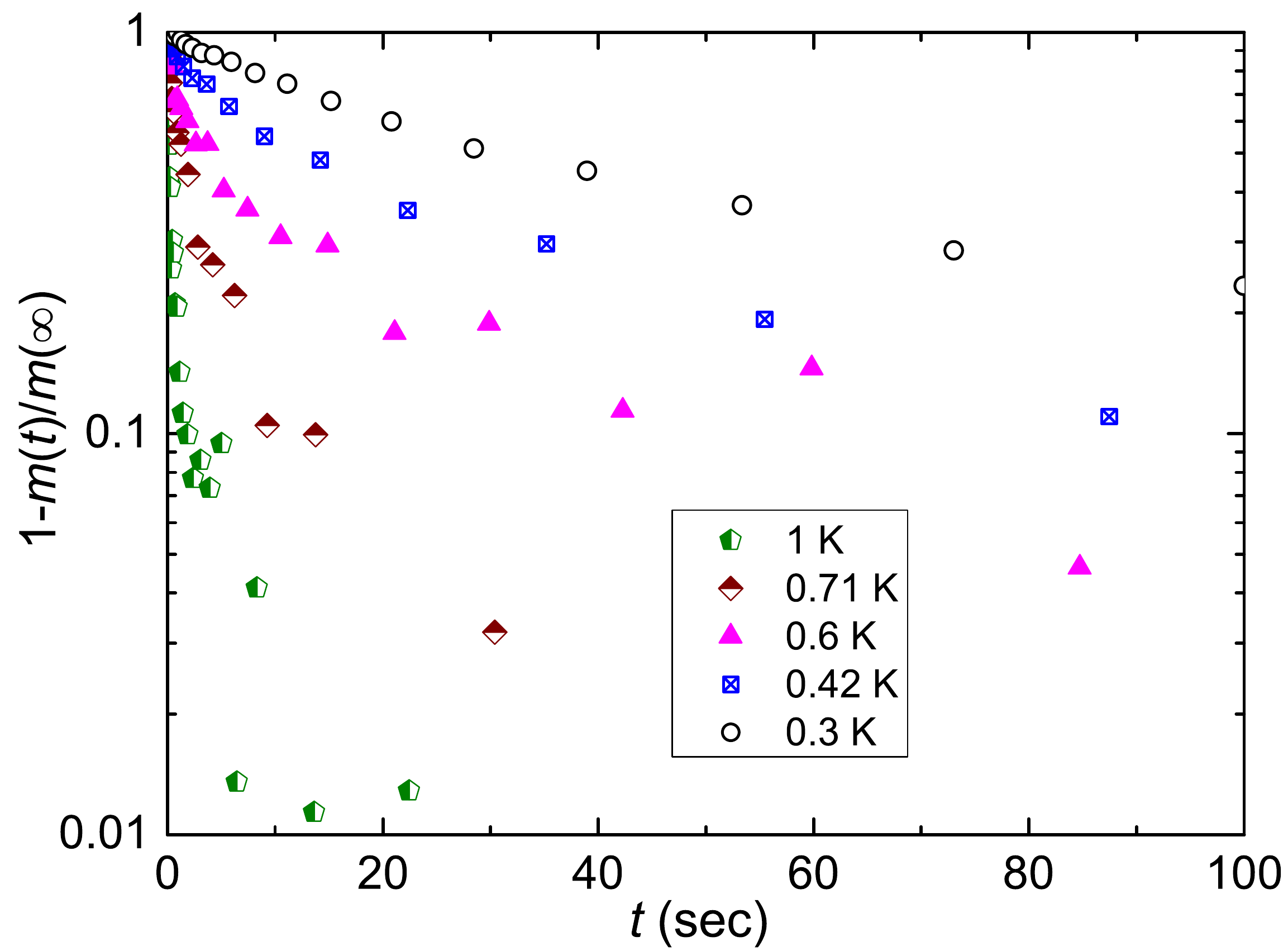} 
\par\end{centering}
\caption{\label{T1 recovery} Recovery of the longitudinal $^{7}$Li nuclear
magnetisation $m(t)$ plotted as $1-m(t)/m(\infty)$ versus the time
delay $t$ at various temperatures. }
\end{figure}

The $^{7}$Li-NMR spin-lattice relaxation rate ($1/T_{1}$) measurements
were performed with the saturation recovery method to study the low-energy
spin dynamics or to probe the $q$-averaged dynamical susceptibility
of Ag$_{3}$LiRu$_{2}$O$_{6}$ in the temperature range \textcolor{black}{$0.3-210$}\,K
at a transmitter frequency of 95\,MHz ($H\simeq57.3$~kOe). The
recovery of the longitudinal $^{7}$Li nuclear magnetisation was monitored
after a saturating pulse sequence and the data in the low temperature
regime are shown in Fig. \ref{T1 recovery}. The recovery of the longitudinal
magnetisation $m(t)$ was fit to (1-$\frac{m(t)}{m(\infty)}=Aexp(-t/T_{1L})+Bexp(-t/T_{1S})$
) where $T_{1L}$, $T_{1S}$ are the long and short components of
the relaxation time and, $A$ and $B$ are constants.\textcolor{black}{{}
The short component ($T_{1S}$) likely corresponds to an initial fast
relaxation associated with spectral diffusion due to incomplete saturation
of the broad line. The long component ($T_{1L}$) is expected to be
the intrinsic constribution.} However, both the long and the short
$T$$_{1}$ components, follow the same qualitative behaviour. The
Fig. \ref{Kiso and T1}(b) illustrates the variation of $1/T_{1L}$
and $1/T_{1S}$ in the $T$ range 0.3-210~K. A gradual decrease is
seen down to 30~K with a broad plateau around $2$\,K which is followed
by a fall-off with a $T^{4}$ power law at lower $T$. The broad maximum
is not due to any spin freezing as the $^{7}$Li NMR line remains
unbroadened all the way from $120$\,K down to $150$\,mK. The $T^{4}$
variation of $1/T_{1}$ below $2$\,K suggests that the spins remain
dynamic and that the excitations are gapless. The data could be fit
to a gapped behaviour with a gap of about $3$\,K but our heat capacity
data show a power-law decrease. Such power law behavior is also seen
in other spin liquid candidate materials \cite{Dey2017,Kumar2016}.


\subsection{$\mu$SR}

As mentioned before, for the GPS experiments, the background signal
can be neglected. For the LTF data (20~mK < T < 17.5~K) we need
to estimate the background signal $Bgd$. At 3 K, we have data from
the GPS as also the LTF beamlines. We then vary $Bgd$ for the LTF
data so that the background subtracted data for LTF (at 3 K) coincides
with the GPS data at 3 K. This yields $Bgd=0.014(1)$. For clarity,
and to directly compare the GPS and LTF data, the raw curves are presented
in terms of polarization, $P(t)$ where : 
\begin{eqnarray}
P(t)=\frac{A(t)-Bgd}{A_{0}-Bgd}
\end{eqnarray}

\begin{figure}
\includegraphics[bb=0bp 0bp 800bp 710bp,scale=0.3]{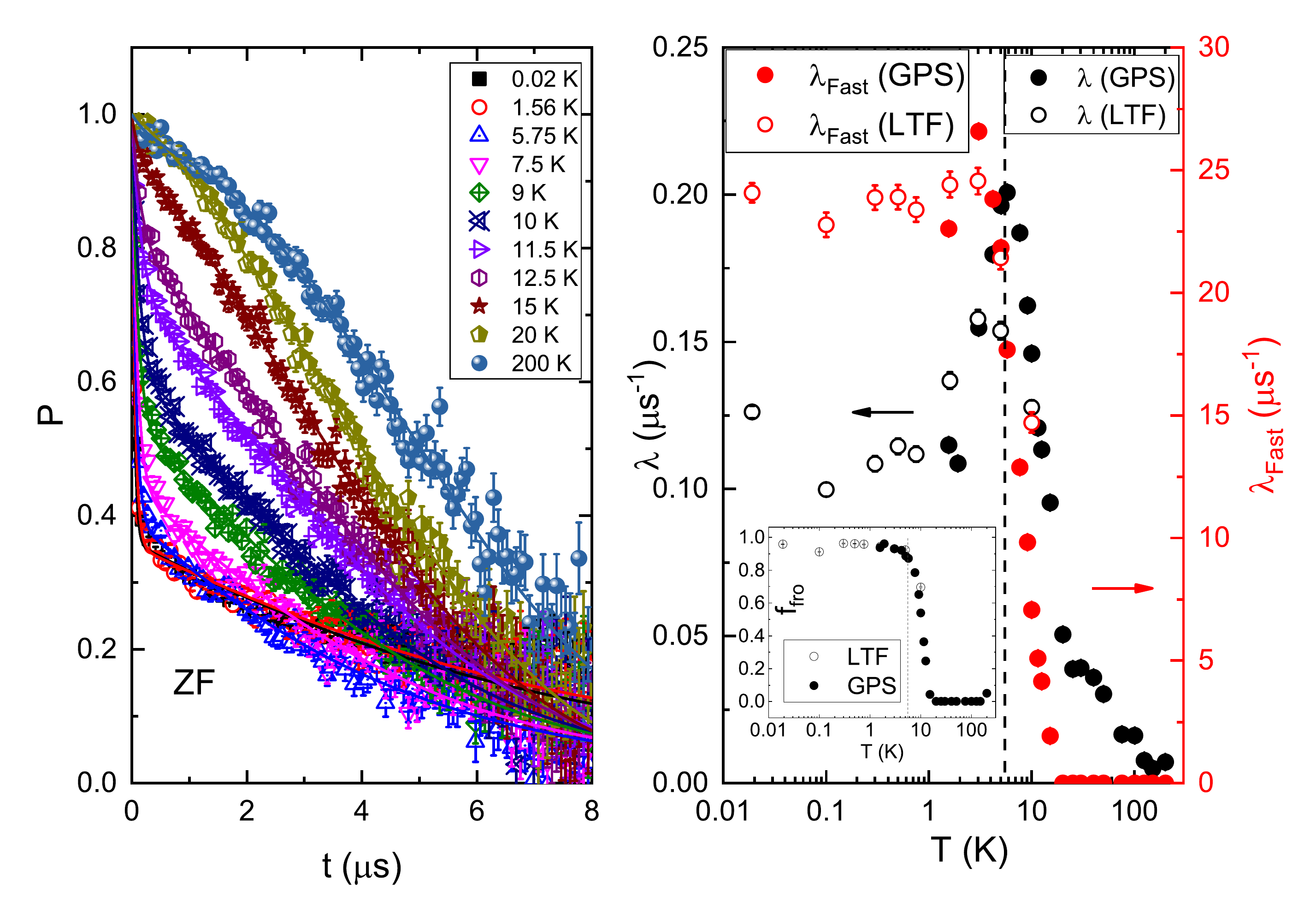} 

\caption{\label{ZFRaw}Left : Polarization versus time at selected temperatures
in zero field. The circles are the raw data and the lines are the
fit with equation \ref{FitZF}. Right : Depolarization rates versus
temperature. The black circles are $\lambda$ (left scale) and the
red ones are $\lambda_{Fast}$ (right scale). Inset : Frozen fraction
versus temperature. The solid circles are the GPS results, the open
circles are the LTF results. The dashed line represents the transition
temperature $T_{g}$}
\end{figure}
We present first the zero field (ZF) experiments that we have performed
from 200~K down to 20~mK (Figure \ref{ZFRaw} left). From these
data, it is clear that there is a transition in the whole compound
between 20~K and 1.56~K. Indeed, at high temperature, in the paramagnetic
regime, the depolarization is quite slow and can be attributed to
the influence of Li magnetic nuclei. We assumed that the depolarization
at 200~K is of a Gaussian form due to static nuclear moments. We
then fitted the 200 K data data with a Kubo-Toyabe function \cite{Hayano1979,Kubo1981}:
\begin{eqnarray}
P_{200~\text{K}}(t)=\frac{1}{3}+\frac{2}{3}\left[1-\left(\sigma_{Nucl}t\right)^{2}\right]e^{-\frac{1}{2}\left(\sigma_{Nucl}t\right)^{2}}
\end{eqnarray}
This yields $\sigma_{Nucl}=0.156(1)$~$\mu$s$^{-1}$ which is directly
linked to the nuclear field $H_{Nucl}$ via $\sigma_{Nucl}=\gamma_{\mu}H_{Nucl}$
where $\gamma_{\mu}=2\pi\times135.5$~$\mu$s.T$^{-1}$ is the gyromagnetic
factor of the muons. We found $H_{Nucl}=1.83(2)$~G which is in the
usual range of nuclear field values.

At low temperatures, the depolarization is rather quick, on the 0.1~$\mu$s
scale, and the depolarization at long time is close to 1/3. This is
characteristic of frozen or quasistatic magnetism. Nevertheless, the
lack of spontaneous oscillations could be directly linked to an absence
of long range ordered magnetism. Therefore, our $\mu$SR experiment
reveals a short range ordered spin-glass-like ground state or a dynamic
ground state with a large distribution of fields. To have better insight
on the ground state probed by $\mu$SR, we fitted the data with an
equation containing two relaxing components which has been used for
other spin-glass systems \cite{Aczel2016,Guo2016} : 
\begin{eqnarray}
 & P(t)=f_{Fro}\left(\frac{2}{3}e^{-\lambda_{Fast}t}+\frac{1}{3}e^{-\lambda t}\right)\label{FitZF}\\
 & +\left(1-f_{Fro}\right)P_{200~\text{K}}e^{-\lambda t}\nonumber 
\end{eqnarray}
where $f_{Fro}$ is the fraction of sample in the frozen state, $\lambda_{Fast}$
accounts for the fast depolarization at short time and represents
the distribution of quasistatic fields in the sample and $\lambda$
accounts for the electronic magnetism and could be linked to its fluctuations.
Note that we also tried to fit the data using the dedicated spin glass
function \cite{Uemura1985} as well as dynamic and static Kubo-Toyabe
functions which resulted in poorer fits than with equation (\ref{FitZF}).

The results are presented in Figure \ref{ZFRaw} right. The small
differences between GPS and LTF likely arise from the difficulty to
fully characterize the $Bgd$ in LTF. The frozen fraction increases
at 10~K to go to close to 1 below 3~K. Therefore, the transition
is not due to an impurity but presents a bulk character. Further,
from the depolarization rate $\lambda$ we can determine the transition
temperature $T_{g}$. Indeed, $\lambda$ possesses a peak around 5.5(5)~K
which is characteristic of a transition to frozen magnetism. Moreover,
there is a very small plateau at 0.11(1)~$\mu$s$^{-1}$ which could
be related to small fluctuations of the magnetism below 1~K and could
be due to a quasistatic order. Further, below this temperature, $\lambda_{Fast}$
presents a plateau around 25(2)~$\mu$s$^{-1}$ which is due to the
distribution of the quasistatic fields. From this value, one can directly
compute the field distribution $\Delta=\lambda_{Fast}/\gamma_{\mu}=290(30)$~G.

\begin{figure}
\centering \includegraphics[scale=0.35]{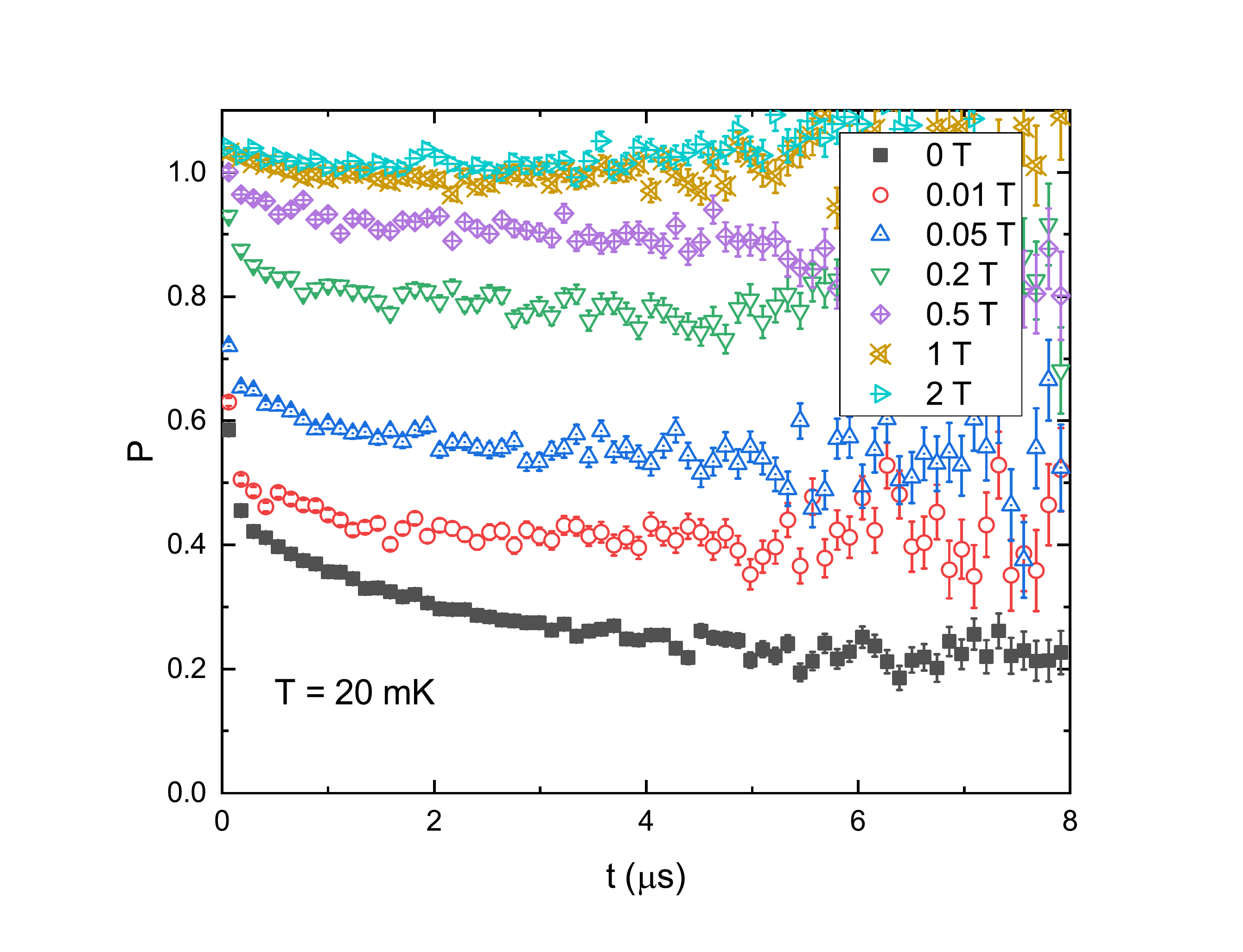}
\caption{\label{Decoup}Muon polarization versus time at 20~mK with different
applied fields.}
\end{figure}
To distinguish between static and dynamic magnetism we applied several
longitudinal fields in the direction of the muons beam. Indeed, in
the case of static magnetism a longitudinal field which is 10 times
the field distribution should decouple the muons whereas in the dynamic
case a longitudinal field 50 times stronger than the field distribution
is needed \cite{Yaouanc2011}. At 20~mK, the muons are almost fully
decoupled under a field of 0.5~T ($\sim17\times\Delta$) (Figure
\ref{Decoup}) indicating that the magnetism observed below 5 K is
at the borderline between static and dynamic behavior. Therefore,
due to the lack of spontaneous oscillations and the decoupling experiment,
Ag$_{3}$LiRu$_{2}$O$_{6}$ presents a spin-glass-like ground state
with a transition temperature $T_{g}$=5.5(5)~K based on the $\mu$SR
analysis.

\subsection{Electronic structure calculations}

In order to obtain further insight into the possible origin of the
observed magnetic behaviour, given the apparent absence of gemetric
frustration, we have carried out electronic structure calculations
using the full-potential linearized augmented plane wave (FP-LAPW)
\cite{Madsen2001,Schwarz2002} plus local orbitals method using the
WIEN2K code \cite{Blaha2001}. Exchange and correlation effects are
treated within generalized gradient approximation (GGA) \cite{Perdew1996}
of Perdew-Burke-Ernzerhof including Hubbard $U$ \cite{Dudarev1998}
and spin-orbit coupling (SOC). The double counting correction in the
GGA+$U$ formalism is taken into account within around mean field
approximation \cite{Czyzyk1994}. The calculations were done with
usual values of $U$ and $J$$_{H}$ \cite{Streltsov2013} chosen
for Ru; $U$ = 3.0 eV and Hund's coupling ($J$$_{H}$) = 0.7 eV.
The calculations were also checked for various other values of $U$.
In order to achieve the convergence of energy eigen values, the kinetic
energy cut off was chosen to be K$_{max}$ R$_{MT}$ = 7.0 where R$_{MT}$
denotes the smallest atomic sphere radius and K$_{max}$ gives the
magnitude of the largest K vector in the plane-wave expansion in the
interstitial region. The Brillouin-Zone integrations were performed
with 8 $\times$ 8 $\times$ 6 k-points mesh. The total energies necessary
for the calculation of symmetric exchange interactions \cite{Xiang2011}
were calculated using density functional theory (DFT) and projector
augmented-wave (PAW) method as encoded in the Vienna ${\it ab}$ ${\it initio}$
simulation package (VASP). The kinetic energy cut off of the plane
wave basis was chosen to be 600 eV and a $\Gamma$ centered 4 $\times$
4 $\times$ 6 k-mesh has been used for Brillouin-Zone (BZ) integration. 

\begin{figure}
\centering{}\includegraphics[bb=0cm 0cm 28cm 22cm,clip,scale=0.28]{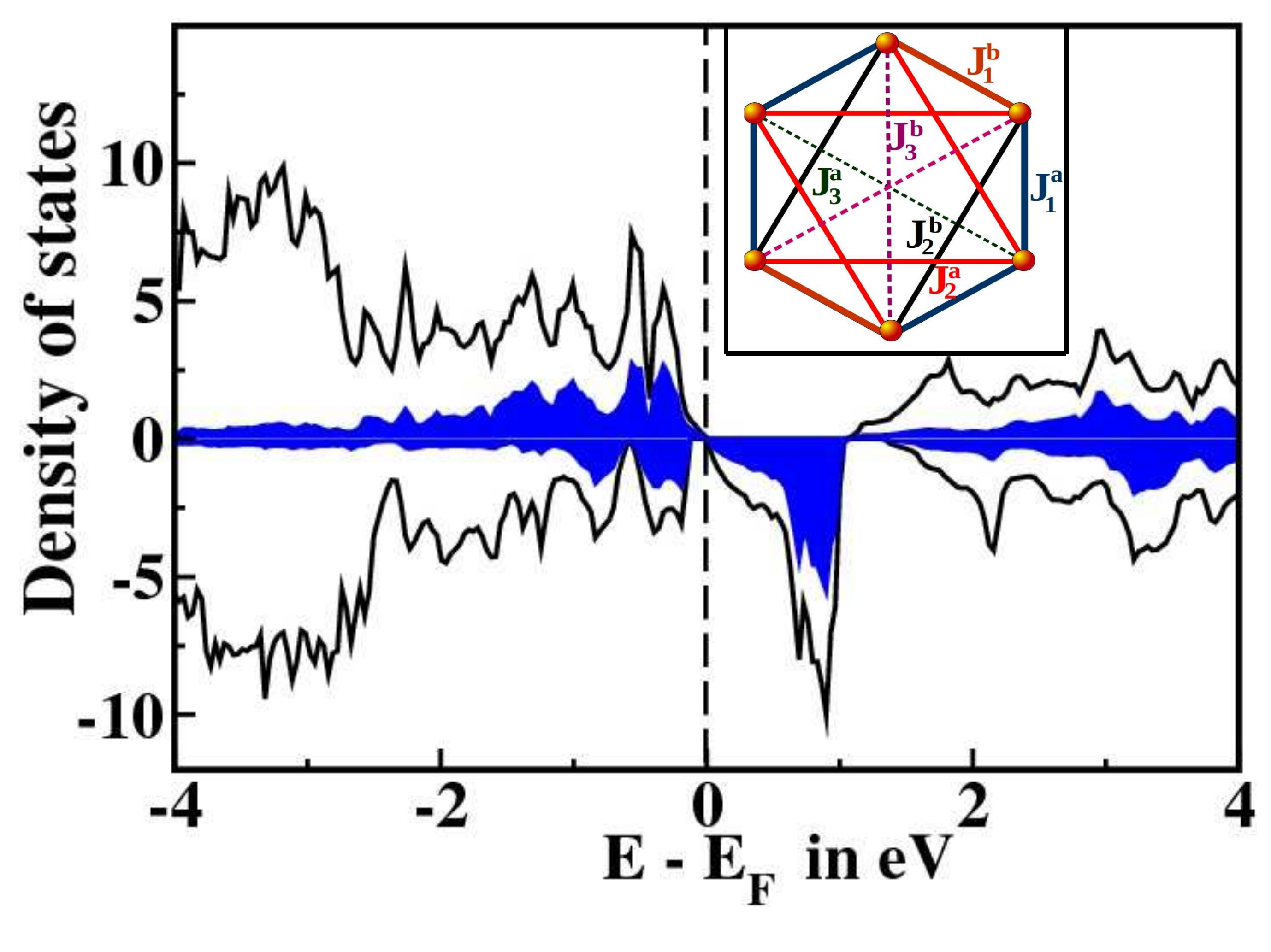}\caption{\label{dos} Spin polarized (ferromagnetic configuration) total and
Ru projected density of states in units of states per eV per formula
unit (blue shaded area) for Ag$_{3}$LiRu$_{2}$O$_{6}$. The exchange
paths and the various exchange interactions are shown in the inset.}
\end{figure}
The spin polarized density of states in the framework of GGA + $U$
with $U=3.0$ eV and $J_{H}=0.7$ eV in the ferromagnetic configuration
\cite{footnote-Ru-honeycomb} is shown in Fig. \ref{dos}. The system
is found to be insulating with the majority Ru $t_{2g}$ spin states
completely occupied while the minority $t_{2g}$ states are only partially
occupied. The $e_{g}$ states for both the spin channels are completely
empty. As a consequence of the monoclinic distortion promoted by Jahn-Teller
active Ru $^{4+}$ ion\textbf{ } the degeneracy of the Ru $t_{2g}$
states is completely lifted which on inclusion of the Hubbard $U$
introduces a gap in the minority spin channel. The total moment is
calculated to be $4\mu_{B}$ (per formula unit) with moment at Ru
and O sites being $1.16\mu_{B}$ and$0.18\mu_{B}$ respectively, suggesting
strong hybridization of the Ru with oxygen states. Next, we have included
spin-orbit coupling in our calculation. The total moment is then calculated
to be 3.97$\mu_{B}$ with spin and orbital moment at the Ru site 1.15$\mu_{B}$
and 0.03$\mu_{B}$, respectively. The small orbital moment suggests
that a spin-only description is valid and neither the \textbf{\textit{LS}}
nor the \textbf{\textit{jj}} coupling schemes should be employed.
These calculations suggest that the system is far away from the $J=0$
limit and $S=1$ description of the system is more appropriate. To
understand the absence of an ordered state, we calculated the first,
second, and third neighbor symmetric exchange interactions mapping
the density functional total energies obtained using Vienna ${\it ab}$
${\it initio}$ simulation package (VASP) within the projector-augmented
wave (PAW) method onto the Heisenberg model following the method proposed
in \cite{Xiang2011}. The magnitude and the sign of the symmetric
exchange interactions are found to be very sensitive to the chosen
configuration for calculation and the size of the simulation cell
suggesting the importance of the higher order magnetic interactions
(biquadratic and four-spin ring couplings) as the usual approximation
of the Hubbard model reducing to the Heisenberg model in the limit
of large $U$ may not be applicable here \cite{Fedorova2015}. The
calculations have been done (with the experimental structural parameters)
for several spin configurations using simulation cells of different
sizes. Our calculations reveal that the nearest neighbor exchange
interactions J$_{1}^{a}$ and J$_{1}^{b}$ are ferromagnetic while
the further neighbor interactions (such as J$_{2}^{a}$, J$_{2}^{b}$,
J$_{3}^{a}$ and J$_{3}^{b}$) are antiferromagnetic. On the other
hand, if we relax the structure and then calculate the couplings,
the nearest and the next-nearest neighbour interations turn out to
be antiferromagnetic. With a larger number of antiferromagnetic further
neighbour couplings, one can still recover a negative $\theta_{CW}$
as observed experimentally. This, coupled with the ring exchange and
biquadratic couplings, introduces frustration in the system and possibly
drives the system away from order.

\section{discussion}

From the wide range of measurements that we have presented together
with first principles electronic structure calculations, let us now
look at things in perspective. Usual bulk susceptibility measurements
did not show any signatures of long-range order down to 1.8 K. Heat
capacity measurements (zero or non-zero applied field) also do not
evidence any peak down to 0.4 K. $^{7}$Li NMR measurements on the
same sample did not show any line broadening nor any peak in the spin
lattice relaxation rate down to 0.15 K. In fact, the $^{7}$Li shift
(which probes the intrinsic susceptibility), starting at room temperature,
increases with decreasing temperature. This suggests that there is
magnetism in the system which must come from Ru$^{4+}$ ions (4$d$$^{4}$).
Note that we tried to fit the $T$-variation of the intrinsic susceptibility
to the formula given by Kotani \cite{Kotani1949} for the Van Vleck
susceptibility of $d^{4}$ systems. This gave rise to very poor fits.
The apparent agreement with the Kotani formula for the susceptibility
of Ag$_{3}$LiRu$_{2}$O$_{6}$ in Ref. \cite{Lu2018} is misleading
as they seem to have considered only the $T$-region where the susceptibility
is nearly constant. Also, our calculations suggest that this is a
spin-only moment. With a continued decrease in temperature, the $^{7}$Li
shift levels off below about 120 K and remains so down to 150 mK.
. The $^{7}$Li NMR linewidth does not show any divergence either
and remains constant down to 150 mK. All the above suggests that the
magnetism in this system somehow gets quenched below 120 K and no
order sets in. Note that the NMR measurements are in a field of about
57 kOe which corresponds to a significant energy scale in our low-$T$
regime (below 5 K or so). Let us now look at the results of low/zero
field measurements. The magnetic heat capacity shows no special features
other than a power law variation at low-$T$ (below about 4 K) and
a broad maximum around 10 K. The low field (25 Oe) ZFC and FC magnetisation
does show some bifurcation at low-$T$ which is sample dependent.
The sample on which we performed the heat capacity and all the NMR
measurements showed a weak ZFC-FC bifurcation (less than 10\%) below
about 3 K. When the same sample was measured in $\mu$SR (zero field),
we found evidence of freezing around 3.5 K. On the other hand, the
sample on which we did detailed $\mu$SR measurements (presented here)
showed a larger ZFC-FC bifurcation and around 7 K. This sample shows
a freezing of moments below about 5.5 K from $\mu$SR. The question
now is how to reconcile the $\mu$SR data with the NMR, susceptibility
and heat capacity data? It appears that the zero-field $\mu$SR and
low-field magnetisation results are consistent with each other. For
the zero field heat capacity, the lack of entropy as well as the broad
peak around 10~K and the power law behavior at temperatures could
perhaps arise in case of a spin glass state. The differing NMR results
might come from a field effect. The $^{7}$Li NMR linewidth should
have shown a critical divergence (or at least a significant increase)
around the freezing temperature. The absence of this suggests that
the applied field prevents the formation of static moments! Nevertheless,
these results are very similar to the one obtained in another honeycomb
compound, Li$_{2}$RhO$_{3}$ \cite{Khuntia2017}. Note also that
the magnetism observed by us in $\mu$SR below 5 K is at the borderline
of static and dynamic. Therefore, even if the $\mu$SR experiments
rule out the possibility of a quantum spin liquid ground state due
to the presence of frozen magnetism below 5.5(5)~K, this compound
does not present a regular spin-glass behavior.

\section{conclusions}

In summary, we have investigated the structural, thermodynamic and
local magnetic properties of a honeycomb structure based novel quantum
material Ag$_{3}$LiRu$_{2}$O$_{6}$ by performing x-ray diffraction,
neutron diffraction, susceptibility, heat capacity and $^{7}$Li-NMR
measurements. The presence of an asymmetric peak in both x-ray and
neutron diffraction profiles is suggestive of a 2D structural ordering
(honeycomb) in the $a-b$ plane. The $\chi(T)$ data infers a strong
antiferromagnetic coupling between the Ru moments without showing
any anomaly down to 2~K, and the neutron diffraction carried out
down to 1.6~K does not detect any magnetic order. Heat capacity displays
a $\sim$ $T^{1.7}$-dependence at low-$T$ and the deduced entropy
change was found to be highly suppressed; $\Delta S\sim11\%$ of that
for an ordered spin-one system. $^{7}$Li-NMR powder spectra measurements
help extracting the instrinsic susceptibility of Ru moments and a
leveling off of the NMR line shift was found for $T$ $\leq$ 120~K.
Our electronic structure calculations provide a clue to the origin
of the observed susceptibility saturation in this system which (in
principle) is not geometrically frustrated. Our study suggests that
the magnetism here is not excitonic in origin. We propose that the
frustration induced by further neighbor couplings and a deviation
from the simple Heisenberg model is responsible for the lack of LRO
in this system. While spin freezing below about 5 K is evidenced from
our zero field $\mu$SR data, from the longitudinal field decoupling
experiments, the moments are at the borderline between static and
dynamic even at 20 mK. It needs to be explored whether defects such
as stacking faults finally drive the system to a frozen state and
whether the pristine system might be a spin-liquid.

We thank Department of Science and Technology (DST), Govt. of India
for financial support through the BRICS project Helimagnets. RK acknowledges
CSIR (India) and IRCC (IIT Bombay) for awarding him research fellowships
to carry out this research work. P. M. Ette acknowledges CSIR, INDIA
for providing financial support under CSIR-SRF Fellowship (grant no.
31/52(14)2k17). AVM thanks the Alexander von Humboldt Foundation for
support. Work at Augsburg was supported by the Deutsche Forschungs
Gemeinschaft (DFG) through the collaborative research center TRR80
(Augsburg/Munich). A.A.Gippius acknowledges the financial support
from the RFBR Grant \textnumero 17-52-80036. We thank Kedar Damle
for useful discussions and Dana Vieweg for technical help.

\bibliographystyle{apsrev4-1}
\bibliography{citation_global}

\begin{thebibliography}{57}
\bibitem [{foo()}]{footnote-Ru-honeycomb}%
  \BibitemOpen
  \href@noop {} {}\bibinfo {note} {The reason for taking a ferromagnetic DOS is
  that it is the simplest magnetic configuration that tells us that the system
  is insulating upon inclusion of a Hubbard U.}\BibitemShut {Stop}%
\end{thebibliography}

\end{document}